\documentclass[12pt]{article}
\usepackage[T1]{fontenc}
\usepackage[utf8]{inputenc}
\usepackage[a4paper,margin=1in]{geometry}
\usepackage{setspace}
\usepackage{amsmath,amssymb,amsthm,bm}
\usepackage{graphicx,xcolor}
\usepackage{booktabs,multirow}
\usepackage{algorithm}
\usepackage[noend]{algpseudocode}
\usepackage{needspace}
\usepackage[bottom]{footmisc}
\usepackage{authblk}
\usepackage[authoryear]{natbib}
\usepackage{xr}
\usepackage[colorlinks,linkcolor=black,citecolor=black,urlcolor=black]{hyperref}

\DeclareMathOperator*{\argmin}{arg\,min}
\DeclareMathOperator*{\argmax}{arg\,max}
\newcommand{\bA}{\mathbf{A}}
\newcommand{\bD}{\mathbf{D}}
\newcommand{\bI}{\mathbf{I}}
\newcommand{\bM}{\mathbf{M}}
\newcommand{\bV}{\mathbf{V}}
\newcommand{\bX}{\mathbf{X}}
\newcommand{\ba}{\mathbf{a}}
\newcommand{\bg}{\mathbf{g}}
\newcommand{\bh}{\mathbf{h}}
\newcommand{\bx}{\mathbf{x}}
\newcommand{\bz}{\mathbf{z}}
\newcommand{\bOmega}{\boldsymbol{\Omega}}
\newcommand{\bSigma}{\boldsymbol{\Sigma}}
\newcommand{\bbeta}{\boldsymbol{\beta}}
\newcommand{\bgamma}{\boldsymbol{\gamma}}
\newcommand{\blambda}{\boldsymbol{\lambda}}
\newcommand{\bmu}{\boldsymbol{\mu}}
\newcommand{\btheta}{\boldsymbol{\theta}}

\numberwithin{equation}{section}
\def\be{\begin{equation}}
\def\ee{\end{equation}}
\def\bea{\begin{eqnarray}} 
\def\eea{\end{eqnarray}}
\def\bd{\begin{displaymath}}
\def\ed{\end{displaymath}}
\def\bda{\begin{eqnarray*}}
\def\eda{\end{eqnarray*}}
\def\bsm{\begin{small}}
\def\esm{\end{small}}

\def\bnt{\begin{enumerate}}
\def\ent{\end{enumerate}} 
\def\bsc{\begin{scriptsize}}
\def\esc{\end{scriptsize}}

\newtheorem{theorem}{Theorem}[section]
\newtheorem{as}{Condition}[section]

\newcommand{\wh}{\widehat}

\makeatletter
\newcommand{\figcaption}{\def\@captype{figure}\caption}
\newcommand{\tabcaption}{\def\@captype{table}\caption}
\makeatother

\newcommand{\ve}{{\varepsilon}}

\newcommand{\wt}{\widetilde}

\newcommand{\p}{\mathbb{P}}

\newcommand{\mI}{{\mathbb I}}

\newcommand{\bzero}{{\mathbf 0}}

\newcommand{\revise}[1]{{\color{black}#1}}

\title{A Latent Semiparametric Mixture Model for Robust Heterogeneity Learning in High Dimensions}
\author[1]{Lyuou Zhang}
\author[2]{Boxiang Wang}
\author[3]{Lulu Wang}
\author[4]{Wen Zhou}
\author[5]{Hui Zou\thanks{Corresponding author: Hui Zou, Department of Applied Mathematics and Statistics, Johns Hopkins University, Baltimore, MD, USA. Email: \href{mailto:hzou20@jh.edu}{hzou20@jh.edu}.}}
\affil[1]{School of Statistics and Data Science, Shanghai University of Finance and Economics, Shanghai, China}
\affil[2]{Department of Statistics and Actuarial Science, University of Iowa, Iowa City, IA, USA}
\affil[3]{Gilead Sciences, Foster City, CA, USA}
\affil[4]{Department of Biostatistics, School of Global Public Health, New York University, New York, NY, USA}
\affil[5]{Department of Applied Mathematics and Statistics, Johns Hopkins University, Baltimore, MD, USA}

\date{}

\begin{document}
\doublespacing
\maketitle

%%%%%%%%%%%%%%%%%%%%%%%%%%%%%%%%%%%%%%%%%%%%%%%%%%%%%%%%%%%%%%%%%%%%%%%%%%%%%%

\begin{abstract}
Learning heterogeneity from high-dimensional data has become a fundamental task across many data-driven fields. Cluster analysis is a widely used unsupervised tool for this purpose, and many existing methods have been adapted to high-dimensional settings through sparsity constraints. In contrast to algorithmic approaches, model-based clustering offers clean theoretical guaranties and computational simplicity, but can be restrictive in practice due to its reliance on specific parametric distributions. We propose a novel latent semiparametric mixture model for clustering, in which the observed data follow a Gaussian mixture after latent coordinate-wise monotone transformations. Furthermore, assuming a sparse optimal clustering rule in the latent space, we develop a new method, {\it CESME}, tailored for high-dimensional clustering. By allowing unspecified transformations, our framework becomes substantially more flexible and robust than traditional model-based approaches. The clustering error of CESME is shown to approach the optimal clustering error at a rate matching that obtained when the true transformations are known.
Extensive simulations and real data analysis confirm that CESME outperforms a broad range of existing high-dimensional clustering methods, including those designed for single-cell RNA sequencing data. The online supplementary file contains technical proofs, additional discussions, and more numerical results. 
\end{abstract}


\section{Introduction}
\label{intro} 
High-dimensional data heterogeneity has received growing attention in fields such as genomics, precision medicine, and social science. Cluster analysis, a widely used unsupervised learning approach, aims to uncover such heterogeneity by partitioning populations into subgroups. Popular methods include $K$-means \citep{hartigan1979algorithm}, which minimizes within-cluster variation; hierarchical clustering, 
% \citep{ward1963hierarchical}, 
which recursively splits data based on group-wise dissimilarity; and spectral clustering \citep{lu2016convex}, which uses the graph Laplacian and its eigen-decomposition to detect heterogeneity. These approaches are largely algorithmic. In contrast, model-based clustering \citep{fraley2002model} adopts a probabilistic framework, modeling data as drawn from a finite mixture of distributions, with each component representing a cluster. A common assumption is that the components follow Gaussian or some specific parametric distributions \citep{mclachlan2019finite}, which simplifies computation and analysis but can be overly restrictive in practice.

\begin{figure}[t]
  \centering
\includegraphics[width=13cm,height=8.5cm]{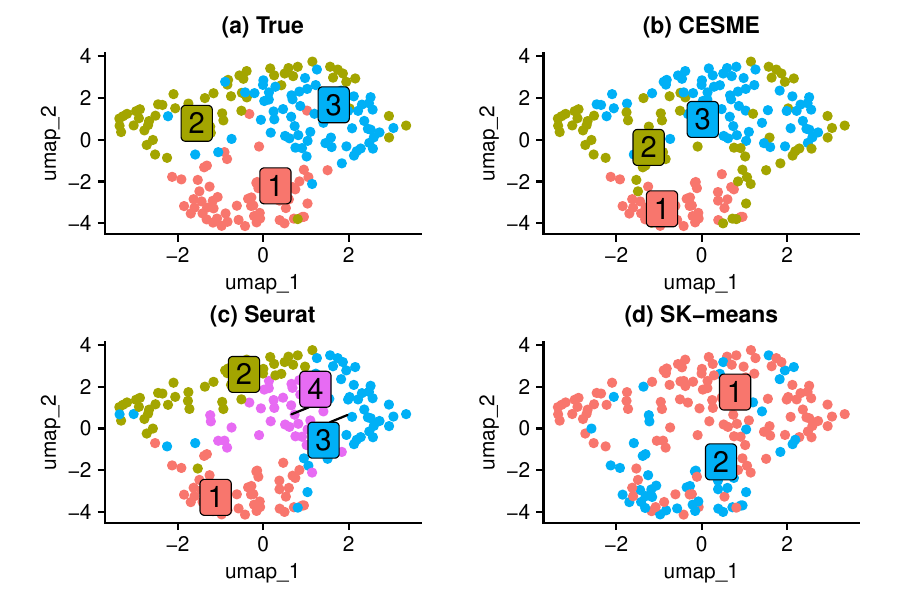} 
    \caption{UMAP plots for clustering the Buettner data, with $182$ cells and $8,989$ genes, showing the true cell labels (a), CESME (b), Seurat (default; c), and sparse $K$-means (SK-means; d). Here, UMAP\_1 and UMAP\_2 denote the coordinates in a two-dimensional embedding that approximately preserves the local structure of the high-dimensional gene expression data.} 
  \label{fig:buettner_umap}
\end{figure}

A key motivation for learning data heterogeneity, involving complex distributions not well captured by simple parametric models, is clustering analysis of single-cell RNA sequencing (scRNA-seq) data, where the primary goal is to uncover biological heterogeneity by identifying unknown cell types \citep{andrews2021tutorial}. Accurate clustering is crucial for reliable downstream analyses, such as detecting highly or spatially variable genes \citep{song2023clusterde, xia2024statistical}. In this paper, we introduce a novel method, {\bf C}lustering  with Lat{\bf E}nt {\bf S}emiparametric {\bf M}ixtur{\bf E} (CESME), which shows strong empirical performance closely aligned with its theoretical guarantees. As a demonstration, Figure~\ref{fig:buettner_umap} presents a two-dimensional Uniform Manifold Approximation and Projection \citep[UMAP;][]{mcinnes2018umap} plot of clustering results on a benchmark scRNA-seq dataset \citep{buettner2015computational}.  
Specifically, we compare results from CESME, Seurat \citep{hao2024dictionary}, and a well-established sparse $K$-means method \citep[SK-means;][]{witten2010framework}, using the true cell labels as reference. CESME clearly outperforms the others, particularly Seurat, a widely used scRNA-seq analysis pipeline, highlighting its effectiveness in clustering  high-dimensional data with complex distributions and feature dependencies.

\begin{figure}[h]
  \centering
   \hspace{-1.6cm}   \includegraphics[width=13.5cm,height=5.25cm]{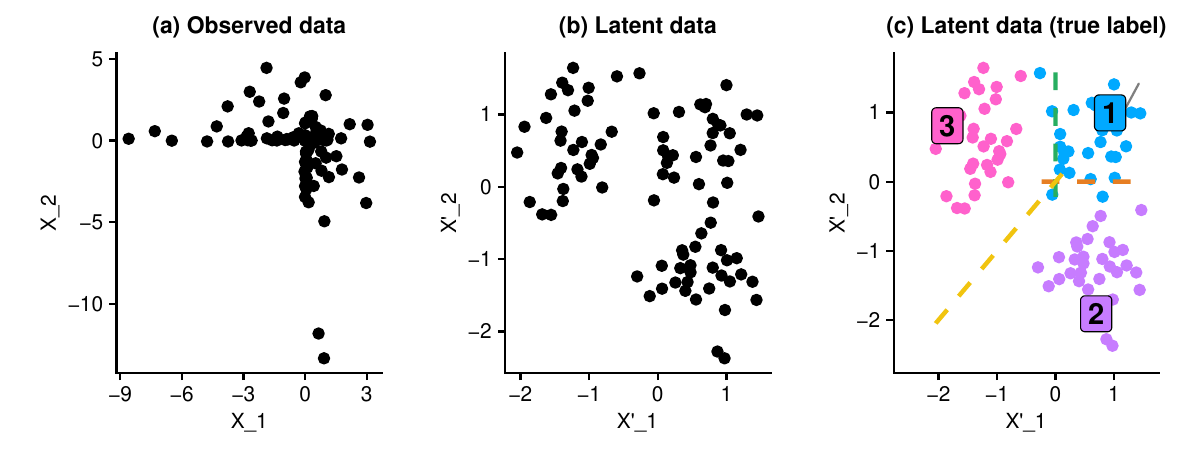} 
    \caption{Scatter plots of simulated two-dimensional data with $90$ observations: (a) observed data; (b) latent data after transformation; (c) latent data with true clustering assignments and the dotted lines indicating decision boundary of the Bayes Rule.} 
  \label{fig:example_simulated} 
\end{figure}

In Figure \ref{fig:example_simulated}, we showcase the key idea behind CESME. Panel (a) displays an observed dataset with $n=90$ observations, where  the clustering boundary is difficult to discern. 
% hard to visualize where to draw
We then apply a coordinate-wise cubic-root transformation to $X_1$ and $X_2$ to obtain the transformed features $X_1'$ and $X_2'$, and plot the latent data based on these new features. As shown in Panel (b), the latent data become more clearly separated, which in turn makes the clustering boundary much easier to visualize in Panel (c). 
 This example illustrates the key idea in CESME: we assume that observed data exhibit clear cluster patterns in the latent space via a coordinate-wise transformation; so a seemingly difficult clustering problem in the observation space is actually much easier in the latent space.

\subsection{Formulation and contributions}
\label{sec11}
To formally introduce CESME, we consider the following data-generating model for the observed data and the clustering problem. First, let $\bz^*=(z^*_1,\ldots,z^*_n)$ denote the cluster membership variables, where for $i=1,\ldots, n$, $z_i^*$'s are independent and identically distributed with $\mathbb{P}(z^*_i=k)=\lambda_k^*$, and $\sum_k \lambda_k^* = 1$. Given $z_i^*=k$, a latent variable $\bX'_i$ is generated from a multivariate normal distribution,  $\bX'_i|z^*_i=k \sim \mathcal{N}(\bmu_k^*, \bSigma^*)$. Let $\bh:=(h_1,\dots, h_p)$ denote a collection of $p$ \revise{strictly} increasing functions. The observed variable $\bX_i$ is obtained via applying $\bh$ coordinate-wise to latent $\bX_i'$, that is, $\bX_i=\bh(\bX'_i)=(h_1(X'_{i1}),\ldots,h_p(X'_{ip}))^{\top}$. For $\ell=1,\ldots,p$, let $g_\ell^*$ denote the inverse of $h_\ell$, and write $\bg^*:=(g_1^*,\dots, g_p^*)$. Then, $X'_{i\ell}=g^*_{\ell} (X_{i\ell})$ for $\ell=1,\ldots,p$.
% the inverse function of $h_\ell$ be $g_\ell^*$ for $\ell=1,\ldots,p$
\revise{Joint identifiability of the latent transformations $\bg^*$ and mixture parameters $\bbeta^*$ remains open for general $K$ and covariance structures. Section B of the supplementary material discusses partial identifiability and establishes identifiability in several structured settings. The clustering guarantee in Theorem \ref{cor:upper_bound} is formulated relative to a fixed data-generating representation: under its stated assumptions, the theorem bounds the excess clustering risk of CESME relative to the corresponding oracle Bayes rule. This guarantee does not require global identifiability and should not be interpreted as establishing uniqueness of the latent parameterization. We also assume that, for some $\delta>0$, $\max_{1\leq\ell\leq p} \mathbb{E}|X_{i\ell}|^{4+\delta} \leq M_{4+\delta}<\infty$, which provides sufficient condition on moment and tail behavior of the observed data $\bX_i$, and enables analyzing the empirical {\it cdf} of $\bX$.}

Note the optimal clustering (Bayes rule) maximizes posterior probability \citep{mclachlan2019finite}.  In our generative model, we see that $\mathbb{P}(z^*_i|\bX_i)=\mathbb{P}(z^*_i|\bX')$, which means that the Bayes rule can be calculated in the latent space. Based on the Gaussian assumption, the posterior maximization reduces to minimizing the quadratic loss $[\{\bg^*(\bX_i)-\bmu_k^*\}^{\top}(\bSigma^*)^{-1}\allowbreak\{\bg^*(\bX_i)-\bmu_k^*\}/2-\log(\lambda_k^*)]$.
Note that we assume a common covariance matrix, which enables a simplified form of the optimal decision rule
\begin{align}\label{eq:zhat0}
\widehat{z}_i=\argmin\nolimits_k[\{\bg^*(\bX_i)-(\bmu_k^*+\bmu_1^*)/2\}^{\top}\bgamma_k^*-\log(\lambda_k^*)],
\end{align} where the {\it discriminant vectors} are $\bgamma_k^*=(\bSigma^*)^{-1}\allowbreak(\bmu_1^*-\bmu_k^*)$ for $k=2,\dots,K$, and $\bgamma_1^*$ is defined to be zero. The linearity of the optimal rule permits an efficient and optimal clustering in high-dimensional spaces when we further assume that each $\bgamma_k^*$ is $s$-sparse, which has been well demonstrated in \cite{cai2019chime}. Another important observation is that we assume the covariance matrices are equal across clusters only in the transformed space. With nonlinear transformations, which is the case of primary interest here, the covariance matrices in the original space can actually be cluster-dependent. This also distinguishes our approach from that of \citet{cai2019chime} (CHIME). It would be of interest to further allow cluster-dependent covariance after transformation, adding more flexibility to our framework. When $p$ is small, this modeling choice does not substantially increase the complexity of the learning algorithm. In high dimensions, however, one would need efficient algorithms to learn the resulting quadratic Bayes rule. Under additional structural assumptions on the covariances, this may be achieved by incorporating large covariance learning algorithms into our procedure, such as the graphical lasso under a sparse precision matrix assumption \citep{Friedmanetal2007}.

One may wonder why not use cluster-dependent transformations to generate the observed data. We prefer cluster-independent transformations for the sake of interpretability, because in \eqref{eq:zhat0}, $X_{i\ell}$ contributes to the optimal clustering if and only if $g_{\ell}^*(X_{i\ell})$ has contributions via the nonzero elements of the discriminant vector. If we choose cluster-dependent transformations, then the optimal clustering compares 
$[\{\bg^*_k(\bX_i)-\bmu_k^*\}^{\top}(\bSigma^*)^{-1}\allowbreak\{\bg^*_k(\bX_i)-\bmu_k^*\}/2-\log(\lambda_k^*)]$. Thus, we lose the clear interpretability and the inference in high dimensions becomes much more complicated to handle.

Based on the above discussions, we prefer cluster-independent covariance matrices and transformations to strike an excellent balance between model flexibility, interpretability, and learning feasibility which is particularly important in high dimensions. Our modeling strategy has been proven to be practically useful in the benchmark data study in Section \ref{sec:real}.
In fact, our method delivers the best clustering results across six scRNA-seq datasets. Our method is expected to be useful in other applications such as imaging, flow cytometry, metabolomics, and financial data, where original features have highly skewed or heavy-tailed distributions.

Our model is also related to the nonparametric mixture models studied by \citet{hall2003nonparametric}. When $\bSigma^*=\bI$, the transformed coordinates $g_1^*(X_{i1}),\dots,g_p^*(X_{ip})$ are conditionally independent given $z^*_i=k$. Thus, the conditional density factorizes as $f_{\bX\mid z^*=k}(x_1,\dots,x_p)
=\prod_{\ell=1}^p f_{X_\ell \mid z^*=k}(x_\ell)$. Hence, our model includes the conditionally independent nonparametric mixture models in \cite{hall2003nonparametric} as a special case. 
However, we believe that it is important to account for within-cluster dependence among features for the success of our method. We therefore do not impose a pre-specified $\bSigma^*$ and rely on a data-driven $\bSigma^*$. The numerical experiments with nonparametric mixture models in Section \ref{sec41} also support our view.

Under the proposed data-generative model, we aim to approximate the optimal rule via the CESME algorithm. We adopt an alternating procedure that iteratively estimates the latent transformations and the parameters of the mixture of Gaussians. Given previous estimates, it alternates between updating mixture parameters via a modified expectation–maximization (EM) and estimating transformations using weighted empirical conditional cumulative distribution functions (CDFs). This strategy is computationally efficient in high dimensions.

To establish theoretical guarantees for CESME, we identify the population solution path, {\it i.e.}, the population-level alternating algorithm, in parallel with the empirical path using data. The estimation error is then decomposed into the {\it numerical error}, representing the gap between the population path and true parameters, and the {\it statistical error}, capturing the deviation between the empirical and population paths. While our analysis is motivated by EM analysis for the Gaussian mixtures \citep{balakrishnan2017statistical}, the presence of latent transformations makes it challenging to establish gradient stability and restricted concavity of the auxiliary $Q$-function with respect to all model parameters, rendering standard EM analysis for the population path inapplicable. To address this, we identify the appropriate parameter spaces for both the mixture parameters and latent transformations, and carefully analyze the Taylor expansion and its remainders over these spaces. 
  
For parameters in traditional Gaussian mixtures, the estimators are Lipschitz, allowing Rademacher-based analysis to directly yield finite sample error bounds between empirical and population paths. In contrast, the estimators of latent transformations in our model are composite functions of empirical CDFs and EM-estimated mixture parameters, making Rademacher complexity analysis intractable. Instead, we develop a fundamentally different framework by bounding the complexity via the Vapnik–Chervonenkis (VC) dimensions of the corresponding function classes. We show that for clustering $n$ subjects with $p$ features, after sufficiently many yet manageable iterations, the estimation error is of order $O_P[\{s\log (np)/n\}^{1/2}]$, with sparsity $s$ defined later. This establishes both convergence and estimation consistency of our algorithm. Moreover, we benchmark CESME's clustering performance against the optimal Bayes rule, and show that they are nearly risk-equivalent, with excess risk of order $O(s\log (np)/n)$. 
\revise{If transformations were known, the excess-risk bound would be $O(s\log (p)/n)$ according to \cite{cai2019chime}. The bound with unknown transformations is $O(s\log(np)/n)$, which has the same order when $\log n=O(\log p)$, including the regime $p\gg n$. Thus, in this regime, estimating the marginal monotone transformations incurs no additional cost in the order of the excess-risk bound.}

\subsection{Related works}
This paper aims to develop a practical, off-the-shelf model-based clustering method for high-dimensional data with complex distributions varying across  dimensions and with feature dependencies. While the classical Gaussian mixture model is widely used \citep{mclachlan2019finite,cai2019chime,mai2021doubly}, extensions include mixtures of parametric heavy-tailed distributions \citep{lin2010robust, lee2013model} and nonparametric mixtures \citep{hall2003nonparametric}. See \cite{RafteryReview2023} for a comprehensive review. Parametric mixtures, like Gaussian ones, are often task-specific and not broadly applicable. Nonparametric mixtures offer greater flexibility, but often assume restrictive conditions, such as within-cluster feature independence, which can degrade empirical performance. These models also tend to struggle in high dimensions.

Several methods have been proposed for high-dimensional clustering. Although spectral clustering is popular for low-dimensional data, \cite{zhang2022leave} showed that it can be suboptimal for clustering under sub-Gaussianity. \cite{wang2021manifold} extended it to high dimensions using $\mathcal{L}_1$-regularized principal components, but, as shown in Section \ref{sec:real}, its performance is limited when real data depart substantially from Gaussianity. \cite{pan2007penalized} proposed a modified model-based approach with diagonal covariance and sparsity-inducing penalties for dimension reduction; however, by ignoring within-cluster dependence, it is less effective in more general settings. 
% Although successful in some settings, it fails in broader applications due to ignoring within-cluster dependence. 
As noted earlier, the widely used sparse $K$-means imposes an $\mathcal{L}_1$ penalty on the cluster means and has been extended to sparse convex clustering \citep{wang2018sparse}, although theoretical guarantees for both methods remain incomplete. \cite{jin2016influential} proposed IF-PCA, which selects informative features using Kolmogorov–Smirnov tests and, in high dimensions, higher criticism, before applying PCA and $K$-means on the principal components. While effective for high-dimensional clustering, IF-PCA's guarantees rely on Gaussianity or sub-Gaussianity. \cite{cai2019chime} introduced CHIME, which exploits sparsity in the optimal clustering rule for Gaussian mixtures and offers theoretical guarantees. However, it is closely tied to the Gaussian assumption and may perform poorly when the data are far from
Gaussian.

\subsection{Organization of the paper}
In Section~\ref{sec:est} we present the details of CESME algorithm, with theoretical analysis given in Section~\ref{sec:theory}. Section \ref{sec:simulation} provides simulations supporting the theoretical results and demonstrating CESME’s empirical performance. In Section \ref{sec:real} we apply CESME to six scRNA-seq datasets and compare it with benchmark pipelines. Technical proofs and additional results appear in the online supplementary file.

\section{An alternating algorithm and the model-based clustering}
\label{sec:est}

Throughout the following algorithm, we treat the number of clusters $K$ as known.
We use nested loops to estimate the transformations $\bg^*$ and mixture parameters $\bbeta^*=(\blambda^*,\bmu^*,\bSigma^*)$, where $\blambda^*=(\lambda_1^*,\dots,\lambda_K^*)^{\top}$ and $\bmu^*=((\bmu_1^*)^{\top},\dots,(\bmu_K^*)^{\top})^{\top}$. 
%We use nested loops to estimate $\bg^*$ and $\bbeta^*$. 
At iteration $t$, the outer loop alternates between updating $\widehat{\bg}^{(t)}$ using $(\widehat{\bbeta}^{(t-1)}, \widehat{\bg}^{(t-1)})$ and updating $\widehat{\bbeta}^{(t)}$ via an EM algorithm using $(\widehat{\bg}^{(t)}, \widehat{\bbeta}^{(t-1)})$, which is the inner loop.

\subsection{Latent transformations via conditional CDF inversion}
\label{sec:est_g}
\revise{To fix a convenient coordinate-wise location-scale convention for the latent representation, we impose}
that $\sum_{k=1}^K \allowbreak\lambda_k^* \bmu_k^* = \bzero$, and $(\sigma_{\ell}^*)^{2} + \sum_{k=1}^K\lambda_k^*(\mu_{k\ell}^*)^{2} = 1$ for each $\ell=1, \cdots, p$. Thus, the marginal means and variances of $\bg^*(\bX_i)$ are zero and one, respectively. Otherwise, we can impose a shift-scale transformation on top of the $\bg^*$ functions to achieve that. 

\revise{
\noindent {\bf Remark 1.} 
The above centering and scaling conditions specify a coordinate-wise location-scale convention, achievable by reparameterization, but they do not establish global identifiability. Identifiability in the general case remains open, but is not required for the clustering guarantee in Theorem \ref{cor:upper_bound}, interpreted as discussed in Section \ref{sec11}. See Section B of the supplementary material for detailed identifiability results.
}

By the monotonicity of $g_{\ell}^*$, we have
$ 
\p ( X_{i\ell} \leq x| z_i^*=k ) = \p \{ g_{\ell}^*(X_{i\ell}) \leq g_{\ell}^*(x) |z_i^*=k \}  = \Phi [ (\sigma_{\ell}^*)^{-1}\{g_{\ell}^*(x) - \mu_{k\ell}^*\} ]
$ for any $x$ in the support of $X_{i\ell}$, with the standard normal CDF $\Phi$. Using $\sum_{k=1}^K \lambda_k^* \mu_{k\ell}^* = 0$, this yields the following natural population representation of $g_{\ell}^*$: 
\begin{equation}
g_{\ell}^*(x)  = \sigma_{\ell}^*\sum\nolimits_{k=1}^K \lambda_k^* \Phi^{-1}\{\mathbb{P}(X_{i\ell}\le x|z_i^*=k)\}.\label{e30}
\end{equation} 
%Hence, 
The above equation motivates estimating the latent transformation as a weighted combination of cluster-specific inverse normal CDF transforms, where each conditional CDF is replaced by its empirical estimate.

Given the update $\widehat{\bbeta}^{(t-1)}$ from the $(t-1)$th iteration of the alternating maximization algorithm, we estimate $g_{\ell}^*$ by replacing $\lambda_k^*$ and $\sigma_{\ell}^*$ in \eqref{e30} with $\widehat{\lambda}_k^{(t-1)}$ and $\wh\sigma_{\ell}^{(t-1)}$ respectively, where $\wh\sigma_{\ell}^{(t-1)} > 0$ is the square root of the $\ell$th diagonal entry of $\wh\bSigma^{(t-1)}$. Then, we can estimate $\mathbb{P}(X_{i\ell}\le x|z_i^*=k)$ by
$\widehat \psi_{\ell k}(x) = \{\sum_{i=1}^n \mI(\widehat z_{i}^{(t-1)} = k)\}^{-1}\sum_{i=1}^n \mI(X_{i\ell}\leq x, \widehat z_{i}^{(t-1)} = k)
$, where $\widehat z_{i}^{(t-1)}$ is the cluster assignment from the $(t-1)$th iteration, with its explicit form given in Section~\ref{npn-est}. %For $t=1$, a provable initialization is detailed in Section \ref{sec42}. 
To sum up, we have the update 
\begin{align}\label{eq:ghat}
\widehat{g}_{\ell}^{(t)}(x) = \wh\sigma_{\ell}^{(t-1)}\sum\nolimits_{k=1}^K \widehat{\lambda}_k^{(t-1)} \Phi^{-1} \{ \widehat\psi_{\ell k}(x)\}.  
\end{align}
Since $\wh\bg^{(t)}$ depends on $\widehat \bz^{(t-1)}$ and $\widehat{\bbeta}^{(t-1)}$, and $\widehat \bz^{(t-1)}$ in turn depends on both $\widehat{\bbeta}^{(t-1)}$ and $\wh\bg^{(t-1)}$, it follows that $\wh\bg^{(t)}$ is ultimately a function of $\widehat{\bbeta}^{(t-1)}$ and $\wh\bg^{(t-1)}$. 

\revise{In practice, as the empirical CDF can be zero or one which should be avoided in the inverse normal CDF, we consider truncating $\widehat\psi_{\ell k}(x)$ as $\widetilde \psi_{\ell k}(x)  =
\widehat \psi_{\ell k}(x) \mathbb{I}\{\delta_n \leq \widehat\psi_{\ell k}(x) \leq 1-\delta_n\} + (1-\delta_n)\mathbb{I}\{\widehat\psi_{\ell k}(x) > 1-\delta_n\}+\delta_n\mathbb{I}\{ \widehat\psi_{\ell k}(x) < \delta_n ~\text{or}~ \sum_{i=1}^n \mI(\widehat z_{i}^{(t-1)} = k) =0\}$ and replacing $\widehat\psi_{\ell k}(x)$ by $\widetilde\psi_{\ell k}(x)$ in \eqref{eq:ghat}, 
where $\delta_n = 1/(4n^{1/4}\sqrt{\pi \log n})$ is similarly set as in \citet{lafferty2012sparse} for estimating the nonparanormal graphical model. }

\subsection{Modified EM iterations}
\label{npn-est}

Given $\widehat{\bg}^{(t)}$ and $\widehat{\bbeta}^{(t -1)}$, the update $\widehat{\bbeta}^{(t)}$ is obtained via an EM algorithm, which serves as the inner loop of our algorithm. Let $\widehat{\bbeta}^{(t, r)}$ denote the update at the $r$th iteration of the inner loop, with $r =  1, 2, \ldots, $ and $\widehat{\bbeta}^{(t, 0)}=\widehat{\bbeta}^{(t -1)}$. In the $E$-step, since ${\bgamma}_k^*$'s are assumed to be $s$-sparse, we estimate them by
\begin{align*}
\widehat{\bgamma}^{(t, r)}_k(\rho)=\argmin\nolimits_{\bgamma\in\mathbb{R}^p}\{\bgamma^{\top}{\widehat{\bSigma}^{(t, r)}}\bgamma/2+\bgamma^{\top} ({\widehat{\bmu}^{(t, r)}}_k-{\widehat{\bmu}^{(t, r)}}_1 )+P_{\rho}(\bgamma)\}
\end{align*}
for $k=2,\dots,K$, and set $\widehat{\bgamma}^{(t, r)}_1(\rho)=\bm{0}$. We use the lasso penalty because it has been used in sparse $K$-means \citep{witten2010framework} and CHIME \citep{cai2019chime}.
This step differs from the canonical EM algorithm for updating the discriminant vectors. 

Given $\widehat{\bgamma}^{(t, r)}_k(\rho)$, the cluster assignment for the $i$th observation is updated as $\widehat{z}_i^{(t, r)}(\rho)=\argmin_k [ \{\widehat{\bg}^{(t)}(\bX_i)- (\widehat{\bmu}_k^{(t, r-1)}+\widehat{\bmu}_1^{(t, r-1)} )/2 \}^{\top}\widehat{\bgamma}_k^{(t, r)}(\rho) -\log(\widehat{\lambda}_k^{(t, r-1)}) ]$ by following the Bayes rule. For each $k$, the posterior probability is updated as
\begin{align}
w^{(t, r)}_k(\mathbf{X}_i,\rho) = \dfrac{\widehat{\lambda}_k^{(t, r-1)}\exp[ -\{\widehat{\bg}^{(t)}(\mathbf{X}_i)- (\widehat{\bmu}^{(t, r-1)}_k+\widehat{\bmu}^{(t, r-1)}_1 )/2 \}^{\top}\widehat{\bgamma}_k^{(t, r-1)}(\rho)]}{\sum_{k=1}^K\widehat{\lambda}_k^{(t, r-1)}\exp[ -\{\widehat{\bg}^{(t)}(\mathbf{X}_i)- (\widehat{\bmu}^{(t, r-1)}_k+\widehat{\bmu}^{(t ,r-1)}_1 )/2 \}^{\top}\widehat{\bgamma}_k^{(t, r-1)}(\rho)]}.
\label{eq:alpha1}
\end{align} 
Then, in the $M$-step, the conditional expectation of the log-likelihood with respect to the distribution of $\mathbf{z}_i|\mathbf{X}_i$, with current estimates $\wh\bg^{(t)}$ and $\widehat{\bbeta}^{(t, r-1)}$, is
$$Q_{\wh\bg^{(t)}}(\bbeta|\widehat{\bbeta}^{(t,r-1)}) = n^{-1}\sum\nolimits_{i=1}^n\sum\nolimits_{k=1}^K w^{(t, r)}_k(\mathbf{X}_i)\{\log({\lambda}_k)+\log [\phi\{\widehat{\bg}^{(t)}(\mathbf{X}_i)|{\bmu}_k,{\bSigma}\} ]\}.$$
Maximizing $Q_{\wh\bg^{(t)}}(\bbeta|\widehat{\bbeta}^{(t,r-1)})$ with respect to $\bbeta$ yields the update $\wh \bbeta^{(t, r)}=\argmax_{\bbeta}Q_{\wh\bg^{(t)}}(\bbeta|\allowbreak \widehat{\bbeta}^{(t,r-1)})$, where
{\allowdisplaybreaks
\footnotesize\begin{align*}
\wh\lambda_k^{(t,r)}&= n^{-1}\sum\nolimits_{i=1}^n w^{(t, r)}_k(\mathbf{X}_i),~
\wh\bmu_k^{(t,r)}= \left\{\sum\nolimits_{i=1}^n w^{(t, r)}_k(\mathbf{X}_i)\right\}^{-1}  \sum\nolimits_{i=1}^nw^{(t, r)}_k(\bX_i)\allowbreak\wh\bg^{(t)}(\bX_i),\\
\wh\bSigma^{(t,r)} &= n^{-1}\sum\nolimits_{i=1}^n\sum\nolimits_{k=1}^K w^{(t, r)}_k(\bX_i) \{\wh\bg^{(t)}(\bX_i)-\wh\bmu_k^{(t,r)}\ \} \{\wh\bg^{(t)}(\bX_i)-\wh\bmu_k^{(t,r)} \}^{\top}.
\end{align*}}

\revise{
For a pre-specified threshold $\ve_0$ and small $\epsilon>0$, we define the relative discrepancy $\operatorname{RD}(\bbeta^{[1]},\bbeta^{[2]})
= \sum\nolimits_{k=1}^K \frac{\|\bmu_{k}^{[1]} - \bmu_{k}^{[2]}\|_2}{\|\bmu_{k}^{[2]}\|_{2} + \epsilon}
+ \frac{ \| \bSigma^{[1]} - \bSigma^{[2]}\|_{2}}{\| \bSigma^{[2]}\|_{2} + \epsilon}
+ \sum\nolimits_{k=1}^K \frac{| \lambda_{k}^{[1]} - \lambda_{k}^{[2]} |}{| \lambda_{k}^{[2]}|+ \epsilon}$
and stopping criterion
\begin{equation}\label{eq:stop0}
\operatorname{RD}(\bbeta^{[1]},\bbeta^{[2]}) < \ve_0 .
\end{equation}
In the $t$th iteration of the outer loop, the EM algorithm alternates between the $E$- and $M$-steps, and terminates at iteration $r$ if $\operatorname{RD}(\wh\bbeta^{(t,r)},\wh\bbeta^{(t,r-1)})< \ve_0$. Once the criterion is met, we set $\wh\bbeta^{(t)}=\wh\bbeta^{(t,r-1)}$ and $\wh\bz^{(t)}=\wh\bz^{(t,r-1)}$. The outer loop terminates at iteration $t$ if $\operatorname{RD}(\wh\bbeta^{(t)},\wh\bbeta^{(t-1)})< \ve_0$. Our procedure is summarized in Algorithm~\ref{alg:cesme}. 
}

\begin{algorithm}[h]
\doublespacing
\caption{High-dimensional CESME}
\label{alg:cesme}
\renewcommand{\thealgorithm}{}
{\fontsize{10}{11}\selectfont
\begin{algorithmic}[1]
{
%\vspace{-0.5cm}
%\alglanguage{pseudocode}

\item[] \textbf{Input}: $p$ by $n$ data matrix $\bX=(\bX_1,\dots,\bX_n)$ and prespecified number of clusters $K$.
 
\State \textbf{Initialize}: $\widehat{\bbeta}^{(0)}=(\wh\blambda^{(0)}, \wh \bmu^{(0)},\wh\bSigma^{(0)})$ and $\wh \bg^{(0)}$ (see Section \ref{sec42} for a provable initialization).
 
\State Compute $\wh\bgamma_k^{(0)}$ and $\wh\bz^{(0)}$.  
\For{$t = 1, 2, \ldots$} \{\textbf{Outer Loop}\}  
\State Compute $\widehat{\bg}^{(t)}$ based on $\widehat \bz^{(t-1)}$ and $\widehat{\bbeta}^{(t-1)}$.  
\State Let $\wh \bbeta^{(t, 0)} = \wh \bbeta^{(t - 1)}$.  
\For{$r = 1, 2, \ldots$} \{\textbf{Inner Loop}\}  
\State The $E$-step: compute $\widehat{\bgamma}^{(t, r)}$, $w^{(t, r)}_{k}$, and $\wh\bz^{(t,r)}$ based on $\widehat \bg^{(t)}$ and $\widehat{\bbeta}^{(t,r-1)}$.   
\State The $M$-step: compute $\wh \bbeta^{(t, r)}$  based on $\widehat{\bgamma}^{(t, r)}$, $\widehat \bg^{(t)}$, and $\widehat{\bbeta}^{(t,r-1)}$.  
\If{condition~\eqref{eq:stop0} is satisfied for $\wh \bbeta^{(t, r)}$ and $\wh \bbeta^{(t, r-1)}$}   
\State Terminate the inner loop and output $\wh \bbeta^{(t)} = \wh \bbeta^{(t, r-1)}$ and $\wh\bz^{(t)} = \wh\bz^{(t, r-1)}$.  
\EndIf   
\EndFor   
\If{condition~\eqref{eq:stop0} is satisfied for $\wh \bbeta^{(t)}$ and $\wh \bbeta^{(t-1)}$}  
\State Terminate the outer loop and output $\wh\bz = \wh\bz^{(t)}$.   
\EndIf   
\EndFor   
\item[] \textbf{Output:}  Clustering assignments $\wh\bz$. }  \vspace{-0.1cm}
\end{algorithmic}
}
%\hspace*{\algorithmicindent} 
\end{algorithm}  

Whenever the stopping criterion \eqref{eq:stop0} is met for $\wh\bbeta^{(T)}$ and $\wh\bbeta^{(T-1)}$ at the $T$th outer iteration, our algorithm terminates with $\wh\bz=\widehat{\mathbf{z}}^{(T)}$ as the estimated cluster assignment, and $\wh\btheta=(\wh\bbeta, \wh\bg)=(\wh\bbeta^{(T)}, \wh\bg^{(T)})$ as the final estimates of the parameters $\bbeta$ of the mixture and transformations $\bg$, respectively. 
 
As a practical alternative to the nested two-loop scheme, one may use a one-step alternating schedule: perform a single EM update for $\bbeta$ followed by one transform update for $\bg$, and repeat. This one-step variant is straightforward to implement and often reduces per-iteration cost, though it typically requires more outer iterations to reach a comparable fit. For numerical comparisons between the one-step EM variant and Algorithm \ref{alg:cesme}, see Section F.1 in the supplementary file.

\subsection{Implementation details}
\label{npn-clustering}
 
\revise{The theoretical analysis uses a penalty parameter $\rho$ of order $\sqrt{\log(np)/n}$; see the proof of Theorem~\ref{cor:upper_bound} in the supplementary file. For an estimate $\widehat{\btheta}(\rho)$ obtained with penalty $\rho$, we define the data-driven BIC-type criterion $\mathrm{Cr}(\rho) = 2l_{K}(\wh\btheta(\rho); \bX) - \{b_K+\widehat{s}(\rho)\}\log(p)$, where $l_{K}(\btheta; \bX) =  \sum_{i=1}^n\log(\sum_{k=1}^{K}\lambda_k \phi(\mathbf{g
}(\mathbf{X}_i)|\bmu_k, \bSigma))$, $b_K=K-1+p(K-1)+p(p-1)/2$ is the baseline mixture parameter count, and $\widehat{s}(\rho)=\sum_{k=2}^K\sum_{j=1}^p\mathcal{I}(\widehat{\gamma}_{kj}(\rho)\ne0)$ is the effective sparsity.

At outer iteration $t\ge1$, we update the penalty as $\rho^{(t)}=\nu\rho^{(t-1)} + C_{\rho}^{(t)} \sqrt{\log(np)/n}$, where $\nu\in(0,1)$ is a fixed numerical decay parameter. At the first inner iteration, we select $C_\rho^{(t)}$ from a finite grid $\mathcal{C}$ of positive multipliers by maximizing $\mathrm{Cr}$ evaluated at the corresponding penalty values and estimates. The selected penalty $\rho^{(t)}$ remains fixed for the remaining inner iterations. Section~4.1 specifies the initial penalty $\rho^{(0)}$, the value of $\nu$, and the candidate grid.}

To enforce the zero mean and unit variance condition in practice, we standardize  $\wh\bg(\bX_1),\dots,\wh\bg(\bX_n)$. Specifically, let ${\rm sd}(\wh{g}_{\ell} ) =  [(n-1)^{-1}\sum_{i=1}^n  \{ \wh{g}_{\ell}(X_{i\ell})-n^{-1}\sum_{i=1}^n\allowbreak \wh{g}_{\ell}(X_{i\ell})\}^2]^{1/2}$, and define the standardized transformation as $\widetilde g_{\ell}(x) = \{\wh{g}_{\ell}(x) - n^{-1}\allowbreak\sum_{i=1}^n \wh{g}_{\ell}(X_{i\ell})\}/\{{\rm sd}(\wh{g}_{\ell} )\allowbreak+\varepsilon_{n,p}\}$, where $\varepsilon_{n,p}=C\sqrt{\log(np)/n}$ with constant $C>0$ independent of $n$ and $p$. This small additive term ensures numerical stability when ${\rm sd}(\wh{g}_{\ell})$ is close to zero. 

We have thus far assumed that the number of clusters $K$ is known. To select $K$ in practice, we run Algorithm \ref{alg:cesme} for each candidate $K$ and compute the corresponding $\mathrm{Cr}_K=\max_{\rho}\mathrm{Cr}(\rho)$ using estimates $\wh\btheta(K)$. The selected number of clusters is the one that maximizes $\mathrm{Cr}_K$: $\wh K=\argmax_K\mathrm{Cr}_K$. This approach is widely used in model-based clustering \citep{RafteryReview2023}. Sections \ref{sec:simulation} and \ref{sec:real} demonstrate that CESME combined with the data-driven BIC-type criterion effectively selects the number of clusters, outperforming competing methods in both simulations and real data applications.
We emphasize that this construction is heuristic and effective in practice; accordingly, we refer to $\mathrm{Cr}_K$ as a data-driven BIC-type criterion.

\section{Statistical Analysis}
\label{sec:theory}
 
\subsection{Conditions and parameter space}

\allowdisplaybreaks
 
To establish statistical guarantees, we first define $\Vert\btheta\Vert_s^2=\Vert\bbeta\Vert_{2,s}^2+\Vert\bg\Vert_{\infty,s}^2$, where $\Vert\bbeta\Vert_{2,s}^2=\Vert\bm{\lambda}\Vert_2^2+\sum_{k=1}^K\Vert\bm{\mu}_k\Vert_{2,s}^2+\Vert\bm{\Sigma}\Vert_{2,s}^2$. For a vector $\ba$, matrix $\bM$, and function $\bg$, define $\Vert\ba\Vert_{2,s}=\sup_{\bx\in\Gamma(s),\Vert\bx\Vert_2=1}|\ba^{\top}\bx|$, $\Vert\bM\Vert_{2,s}=\sup_{\bx\in\Gamma(s),\Vert\bx\Vert_2=1}\Vert\bM\bx\Vert_2$. Here, $\Gamma(s)=\{\ba\in \mathbb{R}^p:2\Vert\ba_{S^c}\Vert_1\le 4\Vert\ba_{S}\Vert_1+3\sqrt{s}\Vert\ba\Vert_2,\allowbreak \mbox{ for } S\subset\{1,\dots,p\} \mbox{ and } |S|=s\}$ denotes a cone of sparse vectors, where $\ba_S$ is the subvector of $\ba$ supported on $S$. 
\revise{For $\bx=(x_1,\ldots,x_p)^{\top}$, write $\bg(\bx)=\{g_1(x_1),\ldots,g_p(x_p)\}^{\top}$. For a deterministic sequence $a_{n,p}\in(0,1/2)$, define $\mathcal{D}_{n,p}=\{\bx\in\mathbb R^p:a_{n,p}\le\Phi(\{g_\ell^*(x_\ell)-\mu_{k\ell}^*\}/\sigma_\ell^*)\le1-a_{n,p}\text{ for each }k=1,\ldots,K\mbox{ and }\ell=1,\ldots,p\}$. The sequence $a_{n,p}$ is chosen so that $\mathcal{D}_{n,p}$ is nonempty; further requirements are specified in Lemma C.4 and the subsequent analysis. For functions $\bg$ bounded on $\mathcal{D}_{n,p}$, define the restricted
supremum seminorms $\Vert\bg\Vert_{\infty}=\max_{1\le\ell\le p}
\sup_{\bx\in\mathcal{D}_{n,p}}|g_\ell(x_\ell)|$ and $\Vert\bg\Vert_{\infty,s}=\sup_{\ba\in\Gamma(s),\Vert\ba\Vert_2=1}\sup_{\bx\in\mathcal{D}_{n,p}}|\ba^{\top}\bg(\bx)|$.}
The norm $\|\cdot\|_{2,s}$ and semi-norm $\Vert\cdot\Vert_{\infty,s}$ will provide refined measures of difference between estimated and true parameters, accounting for the sparsity. Finally, we define $R_{\min}=\min_{1\le j<k\le K}\Vert(\bSigma^*)^{-1/2}\allowbreak(\bmu_k^*-\bmu_j^*)\Vert_{2}$ to quantify the separability between clusters, and impose the following conditions on the model.

\begin{as}
% [Sparsity]
The discriminant vectors $\bgamma_k^*=(\bSigma^*)^{-1}(\bmu_1^*-\bmu_k^*)$ are $s_k$-sparse, where $s_k\le s=o(\sqrt{n/\log(p)})$ for each $k\ge2$.
\label{cond_s}
\end{as}

\begin{as}
The parameter $\bbeta^*$ satisfies $c_{\lambda}\le\lambda_k^*\le 1-c_{\lambda}$ for each $k$, and $c_{\Sigma}^{-1}\le\sigma_{\min}(\bSigma^*)\le\sigma_{\max}(\bSigma^*)\le c_{\Sigma}$, where $c_{\lambda}\in(0,1)$ and $c_{\Sigma}>0$.
Also, we assume that $R_{\min}>4\sqrt{\log_{+}(8M/m)}$,
\revise{where $\log_{+}(x)=\log\{\max(1,x)\}$ for $x\ge0$, $M$ and $m$ are positive quantities defined in Lemmas C.1 and C.2 in the supplementary file.}
\label{cond1}
\end{as}

Assumptions similar to Condition \ref{cond_s} have been considered for discriminant analysis \citep{cai2011a,mai2012a} and high-dimensional clustering \citep{cai2019chime}. The rate of $s$ relative to $n^{1/2}$ has also been used in high-dimensional clustering with sparse cluster means \citep{sun2016simultaneous}, ensuring the consistent estimation of $s$-sparse discriminant vectors \citep{johnstone1994minimax}.

Condition~\ref{cond1} requires a reasonable separation between clusters. Similar assumptions have appeared in \cite{balakrishnan2017statistical} and \cite{cai2019chime}. Notably, there exist constants $\overline{M}$ and $\underline{m}$, independent of $n,p$, and $s$, such that $M\le\overline{M}$ and $m\ge\underline{m}$. \revise{Under the mixing-weight and covariance bounds in Condition~3.2, its separation requirement holds whenever $R_{\min}>4\{\log_{+}(8\overline M/\underline m)\}^{1/2}$. Since $\overline M$ and $\underline m$ are independent of $p$ and $s$, this gives a sufficient separation requirement independent of these dimensions}; see the discussion following Lemma~C.2 in Section~A of the supplement. Since our goal is to benchmark clustering performance against the optimal Bayes rule, this requirement on $R_{\min}$ is milder than those in works that focus on recovering true labels  \citep{zhang2022leave,lyu2022optimal}, where $R_{\min}$ is typically assumed to diverge with $n$ and $p$. Particularly, this condition holds when the non-zero entries of $(\bSigma^*)^{-1}(\bmu_k^*-\bmu_j^*)$ are sufficiently large for $j\ne k$. Also, since our alternating maximization algorithm solves a non-convex problem, initialization is necessary as shown in Algorithm 1; we thus impose the following condition to ensure algorithmic convergence.

\begin{as}\label{cond2}
% [Initialization]
The initializations $\wh\bbeta^{(0)}$ and $\wh\bg^{(0)}$ satisfy $\max\{\Vert\wh\bbeta^{(0)}-\bbeta^*\Vert_{2,s}^2,\Vert\wh\bg^{(0)}-\bg^*\Vert_{\infty,s}^2\}\le R_{\min}^2/16$.
\end{as}

Condition \ref{cond2} only requires the initial estimates to lie within a local basin of attraction whose radius is of order $R_{\min}$, rather than to be very close to the true parameters. Since Condition \ref{cond1} places a lower bound on $R_{\min}$, this requirement does not shrink with $n$ and hence does not require the initialization to be consistent. This makes our initialization condition more attainable in practice than those imposed in prior works \citep{cai2019chime,balakrishnan2017statistical,sun2016simultaneous,lyu2022optimal,zhao2020statistical,wang2023statistical}.
For example, $\Vert\wh\bbeta^{(0)}-\bbeta^*\Vert_{2,s}\lesssim s^{-1}$ is assumed in \cite{cai2019chime}, which becomes more restrictive as $s$ diverges, and 
$\Vert\wh\bmu_k^{(0)} - \bmu_k^*\Vert_2 \le C\max_{k'\ne k} \Vert\wh\bmu_k^{(0)} - \bmu_{k'}^*\Vert_2$ for some constant $C>0$ is assumed 
in \cite{zhao2020statistical} and \cite{wang2023statistical}.

\subsection{Guarantees on the estimations}

The estimation guarantees of Algorithm \ref{alg:cesme}, motivated by the traditional EM analysis, rely on decomposing the estimation error into {\it statistical} and {\it numerical} parts. By identifying the population path $\bbeta^{(t,r)}$ and establishing its local concavity and gradient stability (Lemmas C.1 and C.2 in the supplementary file), we show that the numerical error decays geometrically over iterations. 
Since $\widehat{\bg}^{(t)}$ in \eqref{eq:ghat} is a composite function of empirical CDFs and earlier estimates $\widehat{\bbeta}^{(t)}$, we control the statistical error by bounding the VC dimensions, instead of the intractable Rademacher complexity, of their function classes. In particular, the mixture parameter estimators are characterized by projections onto sparse sets with VC dimension $O(Ks)$, and the complexity of $\widehat{\bg}^{(t)}$ is controlled via Donsker preservation results \citep{chernozhukov2014gaussian}; see Lemma C.5 in the supplementary file. 
These arguments yield a statistical error of order $O(\sqrt{s \log(np)/n})$, while the numerical error decays geometrically and reaches the same order after $T_0$ iterations. Likewise, the number of inner-loop iterations is not fixed a priori, and $R_t$ EM updates in the $t$th outer loop suffice to reduce the optimization error below the statistical error. Consequently, the estimators satisfy
\begin{align*}
\Vert\widehat{\bbeta}^{(T_0)}-{\bbeta^*}\Vert_{2,s}
\lesssim \sqrt{\frac{s\log(np)}n},\qquad
\Vert\widehat{\bg}^{(T_0)}-\bg^*\Vert_{\infty,s}
\lesssim \sqrt{\frac{s\log(np)}n},
\end{align*}
with high probability. See Theorems A.1 and A.2 in the supplementary file.

\subsection{Clustering guarantees}

To study the clustering guarantees of CESME, we first consider the optimal Bayes rule $\widehat{\bz}_{\text{opt}}=(\widehat{z}_{\text{opt},1},\dots,\allowbreak\widehat{z}_{\text{opt},n})^{\top}$, where $\widehat{z}_{\text{opt},i}=\argmin_k[(\bgamma_k^*)^{\top}\{\bg^*(\bX_i)-(\bmu_1^*+\bmu_k^*)/2\}-\log(\lambda_k^*)]$, which leverages oracle information. Following \cite{cai2019chime}, we consider 
%mis-clustering error defined by
%is measured by the permutation-invariant Hamming loss:
%\begin{align}\label{eq:loss}
$\mathcal{L}(G)=\min_{\Pi\in\mathcal{S}_K}\mathbb{E}[\mathbb{I}\{\Pi(G(\bX_{\text{new}}))\ne z_{\text{new}}^*\}],$
%\end{align}
where $G$ is a generic clustering rule, $\mathcal{S}_K$ is the set of all label permutations of $\{1,\dots,K\}$, and $z_{\text{new}}^*$ is the true latent label of new observation $\bX_{\text{new}}$.
% and the expectation is taken with respect to $\bX_{\text{new}}$ conditional on $\bX$.
As noted by \citet{cai2019chime}, benchmarking against $\widehat{\bz}_{\text{opt}}$ allows to evaluate the CESME clustering performance objectively, so we compare $\mathcal{L}(\widehat{\bz})$ and $\mathcal{L}(\widehat{\bz}_{\text{opt}})$.  We prove that CESME is nearly risk-equivalent to the Bayes rule,  as shown in Theorem \ref{cor:upper_bound}.

\begin{theorem}\label{cor:upper_bound}
Under Conditions \ref{cond_s}-\ref{cond2}, let $\rho^{(t)}=O(\sqrt{\log(np)/n})$ and $\kappa=4M\allowbreak\exp(-R_{\min}^2/8)/m$.
Suppose further that Algorithm \ref{alg:cesme} is run for $T_0$ many outer iterations, with $R_t$ being the inner iterations at the $t$th outer iteration. 
Let $\widehat{\bz}^{(T_0)}$ be the clustering output of Algorithm \ref{alg:cesme} and $\widehat{\bz}_{\mathrm{opt}}$ the Bayes clustering rule.

If we let
$$T_0\gtrsim \{-\log(\kappa)\}^{-1}\log_{+}[(\Vert\widehat{\bbeta}^{(0)}-\bbeta^*\Vert_{2,s}+\Vert\widehat{\bg}^{(0)}-\bg^*\Vert_{\infty,s})\allowbreak\sqrt{n/\{s\log(np)\}}],$$ 
and 
$R_t\gtrsim \{-\log(\kappa/2)\}^{-1}\log_{+}[\{\Vert\widehat{\bg}^{(t)}-\bg^*\Vert_{\infty,s}+\sqrt{s\log(np)/n}\}^{-1}\Vert\widehat{\bbeta}^{(t,0)}-\bbeta^*\Vert_{2,s}]$ for each $t=1,\dots,T_0$, where $\log_{+}(x)=\log\{\max(1,x)\}$ for $x\ge0$, 
%where $T_0$ and $R_1,\dots,R_{T_0}$ are positive integers.
then, with probability at least $1-n^{-1}$,
$$\mathcal{L}(\widehat{\bz}^{(T_0)})-\mathcal{L}(\widehat{\bz}_{\text{opt}})\lesssim \frac{s\log(np)}n.$$
\end{theorem}

\revise{Theorem~\ref{cor:upper_bound} uses prescribed iteration numbers $T_0$ and $R_t$ to control finite-iteration errors. Since the requirements on $T_0$ and $R_t$ depend on unknown population quantities, we use the relative-change criterion \eqref{eq:stop0} to terminate both loops in practice. 
Under Conditions~\ref{cond_s}-\ref{cond2}, if $s\log(np)/n\to0$ and the iteration counts satisfy the stated requirements, the excess clustering risk of CESME relative to the oracle Bayes rule converges to zero, without requiring knowledge of the transformations $\bg^*$.}
Theorem~\ref{cor:upper_bound} is directly comparable to Theorem 3.2 in \cite{cai2019chime} where the bound for CHIME is $s\log(p)/n$. Note that if we used the true transformations in CESME, then its error bound would be that of CHIME. For the $p \gg n$ case, our upper bound matches the Gaussian-mixture benchmark rate asymptotically because $\log(np) \le 2 \log(p)$.
Therefore, Theorem~\ref{cor:upper_bound} shows that CESME attains the same rate as its oracle counterpart that has access to the true transformations. This theoretical conclusion is also supported by our simulation results.

\revise{Theorem~3.3 of \citet{cai2019chime} establishes a minimax lower bound of order $s\log(p)/n$ over the Gaussian-mixture parameter class considered there, and CHIME attains this rate. This provides a benchmark for our excess-risk bound. When $\log n=O(\log p)$, Theorem~\ref{cor:upper_bound} gives an excess-risk bound for CESME of the same order, $O(s\log(p)/n)$.}

\subsection{Spectral-based initialization}
\label{sec42}

Initialization plays an important role in fitting mixture models. 
For CHIME under Gaussian mixtures, \citet{cai2019chime} employed the HP algorithm \citep{hardt2015tight} when $K=2$. For two-component Gaussian mixtures,  \citet{wu2019randomly}
showed that a random initialization is sufficient for estimation. 
However, such a strategy becomes unreliable when the number of components exceeds two. 
Indeed, \citet{jin2016local} showed that, for Gaussian mixtures with $K\geq 3$, random initialization can lead the EM algorithm to undesirable local optima with non-negligible probability. Alternatively, the ordered-distance method \citep{jana2024provable} applies to more general Gaussian mixtures, but under a fixed dimension $p$.

Our numerical results suggest that CESME is not overly sensitive to initialization, provided that the initializer is reasonable; see Section~\ref{sec:ini} and Figure~\ref{fig:misclus_initial}. 
Some initialization schemes considered there do not satisfy Condition~\ref{cond2}, yet still perform satisfactorily in practice. 
This is compatible with our theory, since Condition~\ref{cond2} is sufficient rather than necessary. 
Nevertheless, from the practical perspective, it is desired to construct an initialization procedure that satisfies Condition~\ref{cond2} with high probability.

Motivated in part by spectral initialization methods for Plackett-Luce mixtures \citep{nguyen2023efficient}, we consider a spectral-based initialization in Algorithm~\ref{alg:initial}. The key observation is that, in the latent space, $\bg^*(\bX)$ follows a Gaussian mixture, so the Gram matrix $\bg^*(\bX)^{\top}\bg^*(\bX)$ admits a low-rank signal-plus-noise decomposition whose leading eigenspace reflects the cluster structure. Although the observed data $\bX$ are obtained through unknown coordinate-wise monotone transformations, this structure is still approximately preserved when the transformations are sufficiently regular near the component centers. In turn, the leading eigenspace of $\bX^\top\bX$ continues to capture the dominant cluster geometry well enough to yield a suitable initializer, provided that the latent cluster signal is not masked by the effect of the unknown coordinate-wise transformations on the Gram matrix, together with the sampling noise. Since these transformations need not affect all spectral directions equally, we introduce a regularization parameter $\alpha_1>0$ to down-weight weaker and less stable directions, thereby stabilizing the resulting spectral representation.

\begin{algorithm}[ht]
\doublespacing
\alglanguage{pseudocode}
\caption{A Spectral-Based Initialization}
\label{alg:initial}
\renewcommand{\thealgorithm}{}
\hspace*{\algorithmicindent}  
{\fontsize{10}{11}\selectfont
\begin{algorithmic}[1]
%\vspace{-0.7cm}
\item[] \textbf{Input}: $p$-dimensional data $\bX=(\bX_1,\dots,\bX_n)$, preconceived  number of clusters $K$, and the normalization parameter $\alpha_1>0$.

\State Perform eigen-decomposition on $\bA=\bX^{\top}\bX$ to obtain the rank-$K$ approximation $\wh\bV\bD_0^2\wh\bV^{\top}$, where $\bD_0^2$ is the diagonal matrix of top $K$ eigenvalues of $\bA$.
        
\State Normalize $\wh\bD=(\bD_0^2+\alpha_1d_{0,\max}^2\bI)^{-1/2}\bD_0$, where $d_{0,\max}^2$ is the largest eigenvalue.

\State Let $\wh\bz^{(0)}$ be the clustering assignments of $K$-means on $\wh\bD\wh\bV^{\top}$.

\State Let $\wh \lambda_k^{(0)} = n^{-1}\sum_i \mathbb{I}(\wh z_{i}^{(0)}=k)$ and $\wt{\bmu}_k^{(0)}=\{\sum_i \mathbb{I}(\wh z_{i}^{(0)}=k)\}^{-1}\sum_i \mathbb{I}(\wh z_{i}^{(0)}=k)\bX_i$ for each $k$. Given $\wh\bz^{(0)}$ and $\wt\bmu_k^{(0)}$, let $(\wt \sigma_{\ell}^{(0)})^2=n^{-1}\sum_{k=1}^K\sum_{i:\wh z_{i}^{(0)}=k}(X_{i\ell}-\wt\mu_{k\ell}^{(0)})^2$ for each $\ell=1,\dots,p$.  

\State Given $\wh\bz^{(0)}$, $\wh \blambda^{(0)}$, and $\{(\wt \sigma_{\ell}^{(0)})^2\}_{\ell=1}^p$, initialize $\wh \bg^{(0)}$ by \eqref{eq:ghat}, $\wh{\bmu}_k^{(0)}=\{\sum_i \mathbb{I}(\wh z_{i}^{(0)}=k)\}^{-1}\sum_i \mathbb{I}(\wh z_{i}^{(0)}=k)\wh{\bg}^{(0)}(\bX_i)$ for each $k$, and $\wh \bSigma^{(0)}=n^{-1}\sum_{k=1}^K\sum_{i:\wh z_{i}^{(0)}=k}\allowbreak\{\wh{\bg}^{(0)}(\bX_i)-\wh\bmu_k^{(0)}\}\{\wh \bg^{(0)}(\bX_i)-\wh\bmu_k^{(0)}\}^{\top}$. 

\item[] \textbf{Output:}  Initial parameters $\widehat{\bbeta}^{(0)}=(\wh\blambda^{(0)}, \wh \bmu^{(0)},\wh\bSigma^{(0)})$ and $\wh \bg^{(0)}$. 
%\item[] \textbf{Output:}  Clustering assignment $\wh\bz$.
\end{algorithmic}
\vspace{-0.5cm}
} 
\hspace*{\algorithmicindent} 
\end{algorithm}

\revise{Proposition D.1 in Section D.2 of the supplementary file states the initialization guarantee for Algorithm \ref{alg:initial}. Under its stated assumptions, the initial estimators satisfy Condition~\ref{cond2} with high probability, up to a permutation of the component labels. The proof is provided by the argument for Proposition~3.1 of \citet{zhang2024gmixture}, which uses random matrix theory \citep{zhang2022non,abbe2022} and leave-one-out perturbation arguments \citep{fan2018formula} to control estimation of the leading eigenspace of $\bX^\top\bX$ and the subsequent $K$-means assignments and parameter estimates. Section D.2 also discusses how the initialization assumptions relate to the conditions for the subsequent estimation and clustering theory.}

\section{Simulations}
\label{sec:simulation}
%\pagebreak[4]
\allowbreak
\subsection{Clustering accuracy}  
\label{sec41}

In this section, we compare CESME with a comprehensive set of methods, including CHIME, sparse $K$-means (SK-means), IF-PCA, spectral clustering (SC), sparse spectral clustering \citep[SSC;][]{lu2016convex,wang2021manifold}. 
Due to space limitations in the figures, we present the results of other competitors in Section E.1 in the supplementary file, including $K$-means, model-based clustering \citep[Mclust;][]{fraley2002model}, multivariate mixture model-based clustering \citep[Nonpara;][]{benaglia2009like},  sparse convex clustering \citep[SCV;][]{wang2018sparse}, Seurat \citep{hao2024dictionary}, Sincera \citep{guo2015sincera},  significance analysis \citep[SA;][]{grabski2022significance}, and nonparametric mixture model \citep[NPEM;][]{hall2003nonparametric}. 

\revise{For CESME, we use the penalty selection procedure described in Section~\ref{npn-clustering}, with initial penalty $\rho^{(0)}=\sqrt{\log(np)/n}$, numerical decay parameter $\nu=0.8$, and candidate grid $\mathcal{C}=\{2^j,j=-4,\ldots,4\}$.} 
We also include CESME-oracle, which assumes the true transformations $\bg^*$ are known. As a baseline, we report the Bayes rate, which is the accuracy of the optimal Bayes rule under the true model.

\begin{figure}[h!]
  \centering
  %\hspace{0.75cm} 
  \includegraphics[scale=0.4]{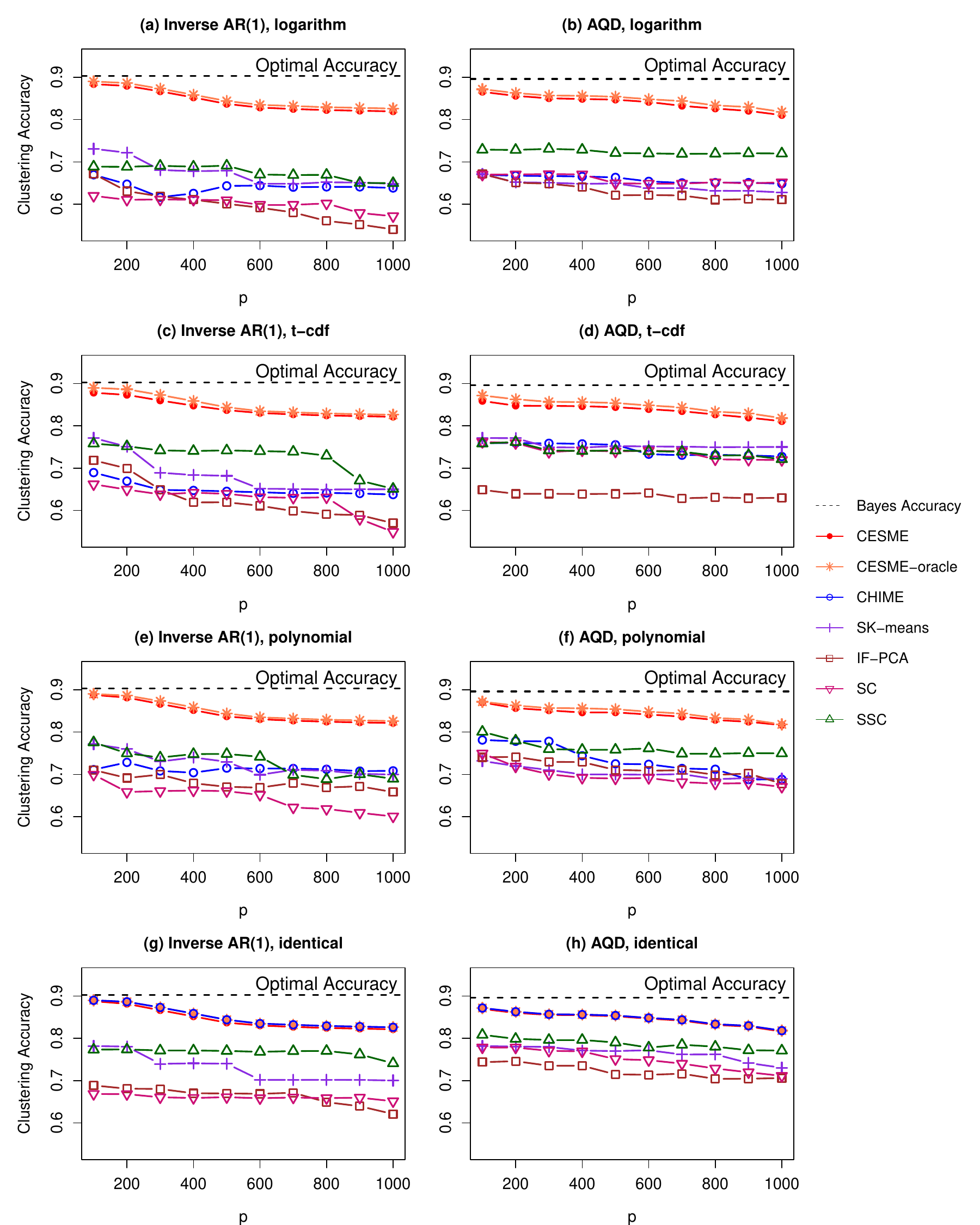}
  \caption{Clustering accuracy comparisons of CESME against CESME-oracle, CHIME, SK-means, IF-PCA, SC, and SSC, with fixed signal-to-noise-ratio $\delta=2.5$ and varying dimension $p$. Covariance structures and transformations are labeled in plot titles. Dashed lines denote the optimal accuracy rates of $0.903$ and $0.896$ for the two covariances, respectively.}
%\vspace{-0.55cm}  
  \label{fig:misclus_K4_ar_strong}
\end{figure}

To generate data $\mathbf{X}_i\in \mathbb{R}^p$ for $i = 1, 2, \ldots, n=400$, we first draw latent variables $\mathbf{Z}_i$ independently from a Gaussian mixture $\sum_{k=1}^K \lambda_k \mathcal{N}(\bmu_k, \bSigma)$, with parameters defined below. For each $\ell$, the samples $Z_{1\ell},\dots,Z_{n\ell}$ are standardized to have zero mean and unit variance. 
% to satisfying Condition \ref{cond0}. 
The observed data are then obtained via component-wise inverse transformations: $X_{i\ell} = g_{\ell}^{-1}(Z_{i\ell})$ for each $i$, where the same transformation $g^{-1}$ is applied across all dimensions. \begin{figure}[h!]
  \centering
  %\hspace{0.75cm} 
  \includegraphics[scale=0.4]{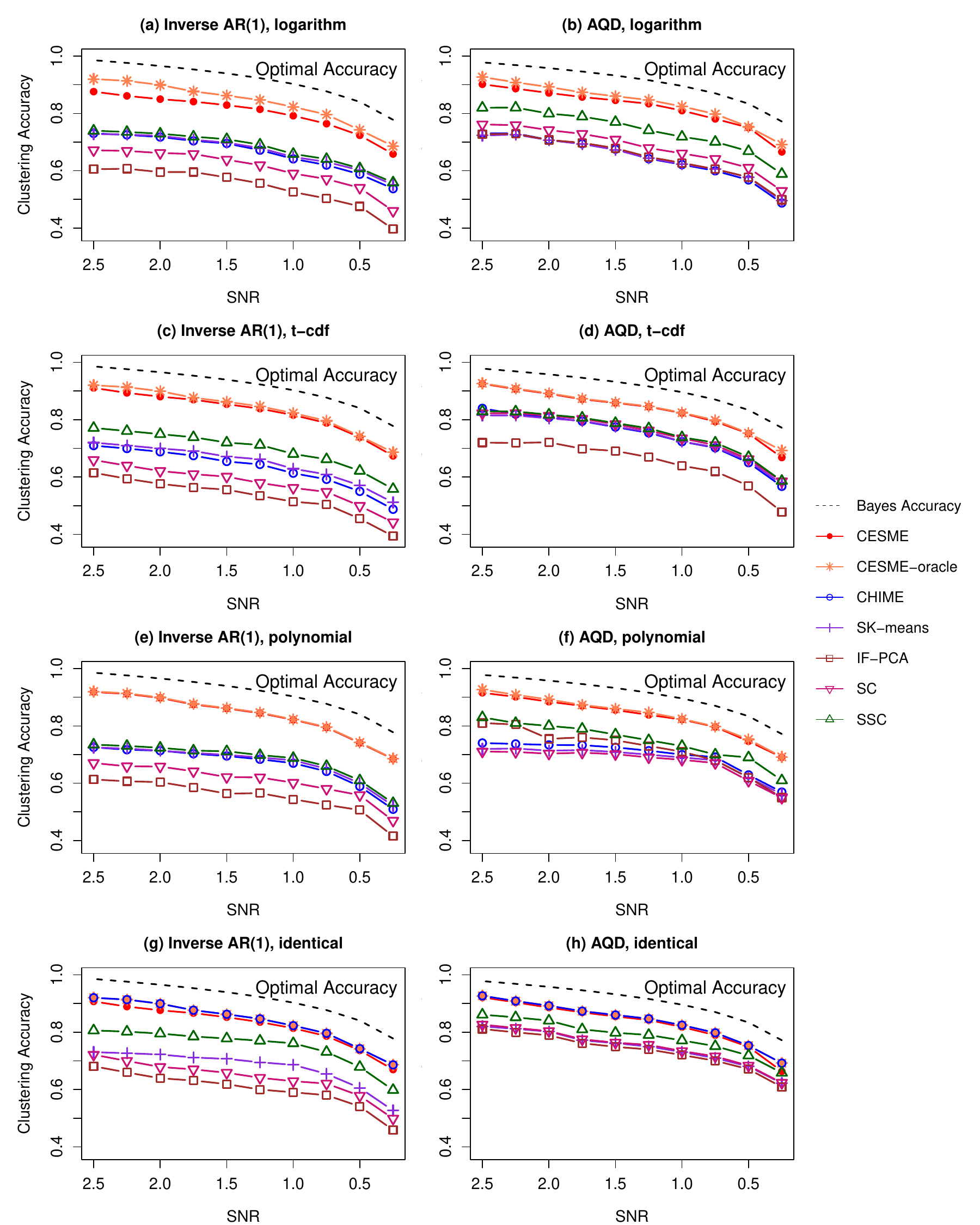}
\caption{Clustering accuracy comparisons of CESME against CESME-oracle, CHIME, SK-means, IF-PCA, SC, and SSC, with fixed dimension $p = 800$ and varying signal-to-noise-ratio $\delta$. Covariance structures and transformations are labeled in plot titles.}
\vspace{-0.5cm}
  \label{fig:misclus_K4_ar_snr_all}
\end{figure}
We consider four transformations: (1) logarithm: $g^{-1}(x)=\exp(x)$; (2) the rescaled inverse $t$-CDF: $g^{-1}(x)=F^{-1}((x-c)/d)$, where $F$ is the CDF of the $t_3$ distribution and $c,d$ rescale $x$ to $(0,1)$; (3) polynomial: $g^{-1}(x) = x^{3}$; and (4) identity: $g^{-1}(x)=x$. We set $K = 4$ clusters with unequal proportions: $\lambda_1 = 0.2$, $\lambda_2 = 0.23$, $\lambda_3 = 0.27$, and $\lambda_4 = 0.3$. Two covariance structures are considered: the inverse AR (1) model, $\bSigma = \bOmega^{-1}$ with $\Omega_{\ell \ell'} = 0.8^{|\ell-\ell'|}$; and the asymptotic quadratic decay (AQD) model with $\sigma_{\ell \ell'}=0.7/(0.7 + |\ell-\ell'|)^2$. For each $\bSigma$, the cluster means are defined as $\bmu_1=\bSigma(\bgamma_1+\bgamma_2+\bgamma_3)$, $\bmu_2=-\bSigma\bgamma_1$, $\bmu_3=-\bSigma\bgamma_2$, and $\bmu_4=-\bSigma\bgamma_3$, where $\bgamma_1$, $\bgamma_2$, and $\bgamma_3$ are sparse signal vectors:
$\bgamma_1 = 4\delta (1,\dots,1,0,\dots,0)^{\top}$ with nonzero entries in positions $1$-$10$; $\bgamma_2 = 4\delta  (0,\dots,0,1,\dots,1,0,\dots,0)^{\top}$ in positions $11$ to $20$; and $\bgamma_3$ similarly in positions $21$ to $30$. The parameter $\delta$ controls signal strength and thus governs cluster separability. The theoretical conditions for initialization with Algorithm 2 are satisfied under both covariance settings, across all signal strengths and dimensions.
%, making our spectral-based initialization in Algorithm \ref{alg:initial} applicable. 
For simulations, the true value of $K$ is provided to all competing methods.

Figures \ref{fig:misclus_K4_ar_strong} and \ref{fig:misclus_K4_ar_snr_all} display clustering accuracies for  $\delta = 2.5$ with $p$ ranging from $100$ to $1000$, and for varying  $\delta\in [0.25,2.5]$ with fixed $p=800$, respectively. CESME consistently outperforms all competing methods under the three non-identity transformations. Although CHIME and SSC generally perform better than the other methods, they still fall notably short of CESME.
The gap between CESME and CESME-oracle is negligible, suggesting that the transformation estimation \eqref{eq:ghat} is satisfactory and robust, even with complex covariances and unbalanced clusters. When the transformation is identity, CESME performs nearly identically to CHIME, which in this case aligns with CESME-oracle. The numerical results support the theory in Section \ref{sec:theory}. 
The comparison of CESME with additional competitors is presented in Figures S.3 and S.4 in Section E of the supplementary file. It is observed that $K$-means, Mclust, Nonpara, and NPEM perform worse than the competitors displayed in Figures \ref{fig:misclus_K4_ar_strong} and \ref{fig:misclus_K4_ar_snr_all} even under identity transformations, since they do not exploit model sparsity. 
CESME continues to dominate these competitors.

\begin{table}
\def\arraystretch{1}%
\caption{\revise{Frequency of correctly selecting the number of clusters $K$ by maximizing the data-driven BIC-type criterion $\text{Cr}_K$ or using the gap statistic, based on CESME with $n=400$, $p=800$, and $\delta=2$.}}
\centering
\scalebox{1}{ 
\begin{tabular}{llccccc}
\toprule
&Covariance & logarithm & $t$-CDF & polynomial & identical & combination \\ 
  \midrule
\multirow{2}{1.5em}{Cr} & inverse AR (1) & 0.88 & 0.88 & 0.87 & 0.90 & 0.82 \\ 
 &  AQD & 0.85 & 0.88 & 0.84 & 0.92 & 0.79 \\ 
   \midrule
\multirow{2}{1.5em}{Gap} & inverse AR (1) & 0.58 & 0.61 & 0.59 & 0.67 & 0.55 \\ 
 &  AQD & 0.50 & 0.55 & 0.52 & 0.59 & 0.51 \\ 
   \bottomrule
\end{tabular}
}
% \vspace{-.5cm}
\label{tab:bic_largep_smallsnr}
\end{table}

 \subsection{On the selection of $K$ via the data-driven BIC-type criterion}
To assess CESME's ability to select $K$ using the data-driven BIC-type criterion $\mathrm{Cr}_K$, we used the same data generation models with an additional setting involving heterogeneous transformations across dimensions (denoted “combination”; see Section E of the supplementary file). For comparison, we included the gap statistics \citep{tibshirani2001estimating} which is a very popular technique for determining the number of clusters. Simulations were conducted with $n=400$, $K=4$, $p=100,200,\dots,1000$ and $\delta=0.25,0.5,\dots,2.5$. For each setting, CESME was run with $K\in\{2,\ldots, 8\}$, and the value maximizing $\mathrm{Cr}_K$ was selected. Results based on $100$ replicates are reported in Table 1. Overall, the proposed data-driven BIC-type criterion ($\mathrm{Cr}_K$) achieves strong accuracy in selecting the correct $K$, substantially outperforming the gap statistic.  

\begin{figure}[ht]
  \centering 
  \includegraphics[scale=0.45]{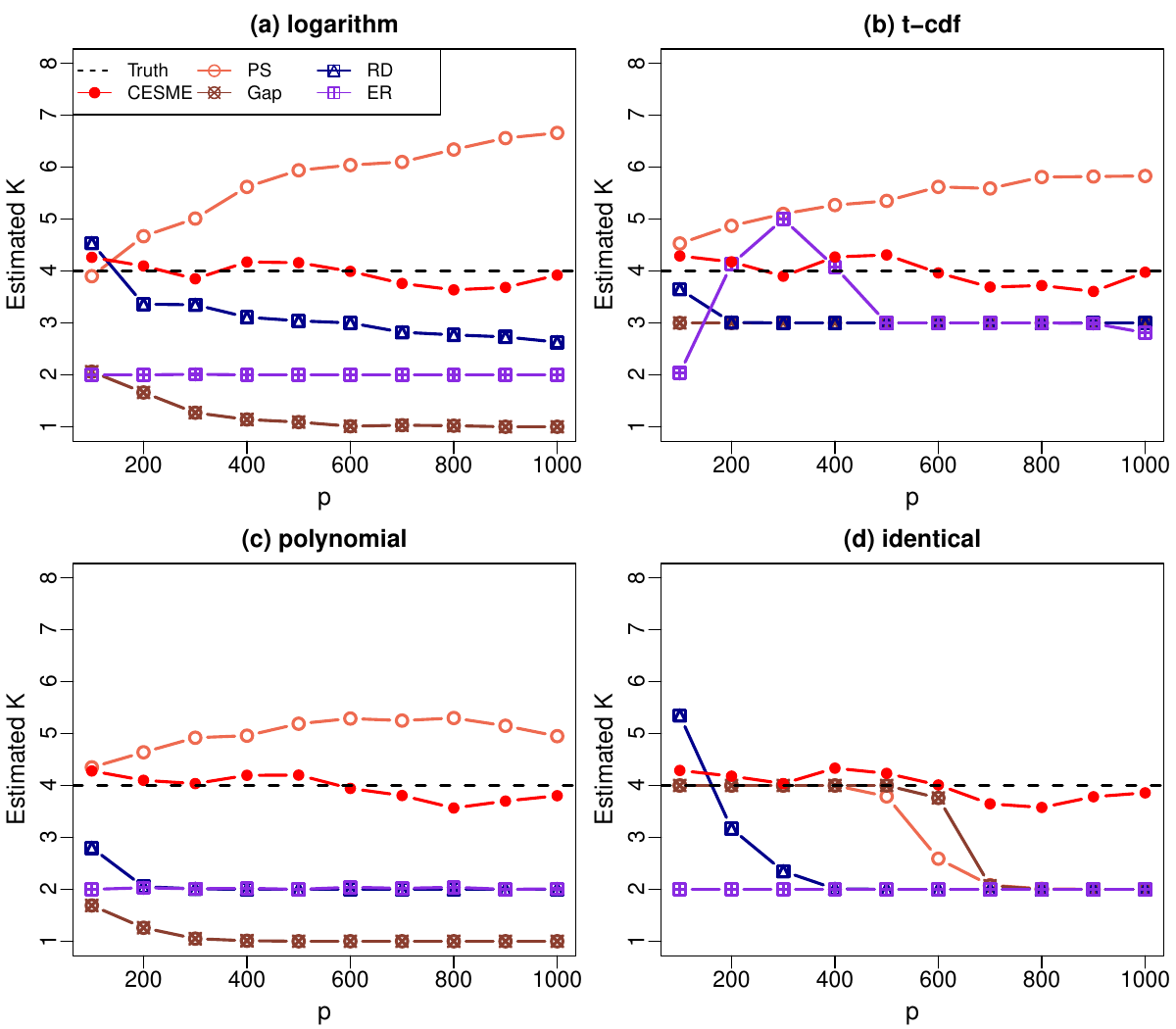}
  \caption{Estimated number of clusters for AQD when $K=4$, $\delta=2.5$, and $p$ varies from $100$ to $1000$. The dashed line shows the true $K=4$.}
  \label{fig:chooseK_K4_frac_TRUE_p_response}
\end{figure}

We further compared our proposed data-driven BIC-type criterion with several additional methods for selecting $K$, including prediction strength \citep[PS;][]{tibshirani2005cluster}, gap statistic \citep[GAP;][]{tibshirani2001estimating}, rank detection \citep[RD;][]{jin2016influential}, and the eigenvalue-ratio test (ER). Figure \ref{fig:chooseK_K4_frac_TRUE_p_response} summarizes the results. Overall, the proposed data-driven BIC-type criterion outperforms the competing methods.

\subsection{Impact of initialization}
\label{sec:ini}

\begin{figure}[h]
  \centering
  \includegraphics[scale=0.4]{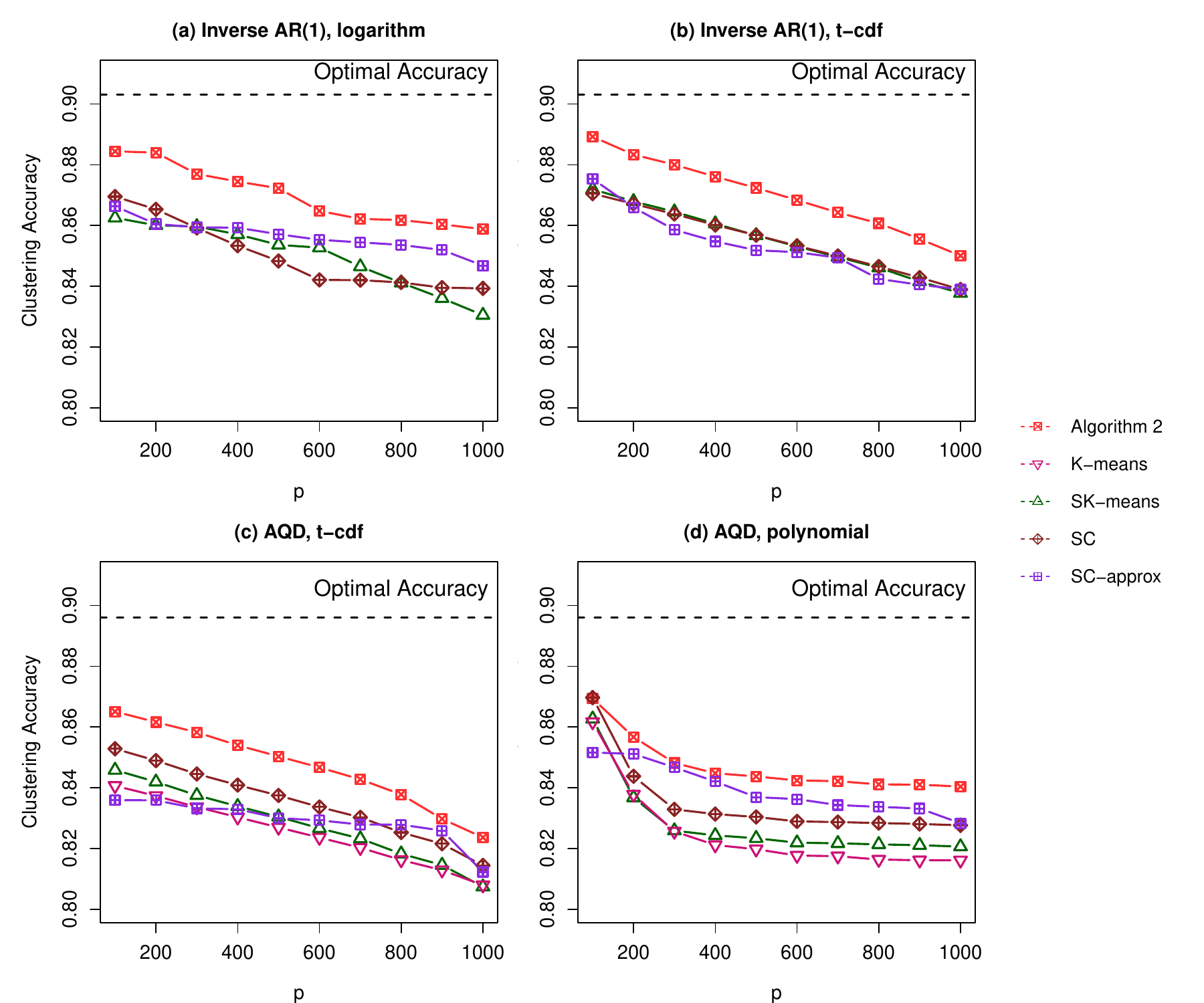}
  \caption{Comparisons of the clustering accuracy for CESME under different initializations. The signal-to-noise-ratio parameter $\delta=2.5$ is fixed and the dimension $p$ is varied. The covariance structure and transformation are specified in figure titles. The dashed lines represent the optimal accuracy rates of $0.903$ and $0.896$ for the two covariance structures, respectively.}
  \label{fig:misclus_initial}
\end{figure}

Furthermore, we conduct numerical experiments to examine the performance of CESME under different initialization strategies with the same simulation settings in Section~\ref{sec:simulation}. 
For each setting, we run CESME initialized by Algorithm \ref{alg:initial}, { $K$-means}, sparse $K$-means (SK-means), spectral clustering (SC), and spectral clustering with the $(1 +\epsilon)$-approximate { $K$-means} algorithm (SC-approx). Since the initialization methods except Algorithm \ref{alg:initial} require the true number of clusters $K$ to provide valid initialization, we employ the true $K$ for this simulation. As shown in Figure \ref{fig:misclus_initial}, 
CESME is insensitive to initialization as long as the initialization is reasonable. The proposed initialization, namely Algorithm~\ref{alg:initial} is consistently slightly better than other initialization schemes.

\section{Real Data Analysis}
\label{sec:real}

To demonstrate the performance of CESME, we apply it to six high-dimensional scRNA-seq benchmark data sets from \cite{buettner2015computational,deng2014single,schlitzer2015identification,pollen2014low,ting2014single,treutlein2014reconstructing}, which 
have also been used in prior studies benchmarking clustering methods in scRNA-seq analysis \citep{wang2021manifold, park2018spectral}. Each dataset is preprocessed for normalization and includes ground truth cell annotations. Although clustering is performed in an unsupervised manner, the true labels are used post hoc to evaluate clustering accuracy.

For a comprehensive comparison, we include several methods that performed competitively in the simulation studies: CHIME, $K$-means, sparse $K$-means (SK-means), IF-PCA, spectral clustering (SC), and sparse spectral clustering (SSC). We exclude multivariate mixture model-based methods and sparse convex clustering due to their instability in high-dimensional settings. We also compare CESME with three widely used pipelines, as noted in the Introduction, for a more scientifically relevant evaluation: significance analysis (SA), Sincera, and Seurat. \revise{To determine the number of clusters $K$, we use the data-driven BIC-type criterion proposed in Section \ref{npn-clustering}, evaluated over $K=2,\ldots,12$, for both CESME and CHIME.} For SSC,  
we apply the eigenvalue-ratio test, selecting $K=\argmax_{k>1}\sigma_{k}(\bX^{\top}\bX)/\sigma_{k+1}(\bX^{\top}\bX)$, where $\sigma_k(\bX^{\top}\bX)$ denotes the $k$th largest eigenvalue of $\bX^{\top}\bX$. For $K$-means and SK-means, we adopt the prediction strength test and set $K$ as the largest number of clusters for which the prediction strength exceeds $0.8$ \citep{tibshirani2005cluster}. For IF-PCA, we follow its originally proposed rank detection method.
% \citep{kritchman2008determining}. 
For SA, Sincera, and Seurat, we use their default pipelines to select $K$.

\begin{table}
\def\arraystretch{1}%
\caption{Clustering accuracy (in $\%$) and estimated cluster numbers (in parentheses) on benchmark datasets, with highest accuracies in bold.}
\centering
\setlength{\tabcolsep}{2pt}
\renewcommand{\arraystretch}{1.15}
\resizebox{\linewidth}{!}{%
\begin{tabular}{@{}l*{12}{c}@{}}
\toprule
Dataset & $n,K$ & $p$ & CESME & Seurat & IF-PCA & SA & $K$-means & SK-means & CHIME & SC & Sincera & SSC \\
\midrule
Schlitzer & $251,3$ & $11835$ & \textbf{83.9} (3) & 53.8 (4) & 59.0 (2) & 63.3 (2) & 64.5 (2) & 66.1 (2) & 47.1 (5) & 57.8 (4) & 30.7 (5) & 41.4 (4) \\
Pollen    & $249,11$ & $14806$ & \textbf{79.1} (9) & 77.9 (7) & 66.7 (6) & 70.8 (9) & 27.3 (2) & 27.3 (2) & 57.4 (2) & 27.3 (2) & 52.4 (6) & 18.5 (2) \\
Deng      & $135,7$ & $12549$ & \textbf{73.3} (7) & 73.3 (4) & 52.6 (3) & 46.7 (2) & 45.2 (2) & 45.2 (2) & 50.2 (5) & 50.4 (9) & 28.9 (3) & 29.6 (9) \\
Ting      & $114,5$ & $14406$ & \textbf{79.1} (5) & 78.9 (3) & 64.9 (3) & 68.9 (5) & 62.3 (2) & 65.8 (2) & 26.7 (2) & 57.9 (2) & 32.5 (4) & 22.8 (2) \\
Treutlein & $80,5$ & $959$ & \textbf{68.5} (5) & 66.2 (2) & 60.0 (2) & 60.3 (2) & 66.2 (2) & 65.0 (2) & 61.6 (2) & 45.0 (11) & 36.2 (3) & 30.0 (11) \\
Buettner  & $182,3$ & $8989$ & \textbf{68.1} (3) & 67.6 (4) & 64.8 (2) & 44.5 (2) & 64.8 (2) & 60.4 (2) & 55.8 (2) & 50.0 (4) & 25.3 (6) & 40.7 (4) \\
\bottomrule
\end{tabular}
}
\end{table}

As shown in Table 2, CESME consistently achieves the highest clustering accuracy on five of the six datasets and most accurately estimates $K$, including the challenging Pollen dataset with $11$ true cell clusters, one of which contains only seven cell samples. Identifying $K$ in such cases is difficult, as small clusters tend to be absorbed into larger ones, leading to underestimation. On the Deng dataset, CESME performs comparably to the widely used benchmark, Seurat. The estimated sparsity levels by CESME are $325,163,62,117,53$, and $230$, across the datasets, suggesting that scRNA-seq data exhibit underlying sparse structures. 
%As a result, methods that leverage sparsity, such as CHIME, SK-means, and IF-PCA, also perform reasonably. 
It is worth noting that for datasets with a large number of true clusters, random clustering would result in very low accuracy, and methods that fail to accurately estimate $K$ also tend to perform poorly. Furthermore, substantial non-Gaussianity, as evidenced by gene-wise kurtosis and skewness in Figure S.12 in the supplementary file, makes accurate selection of $K$  even more difficult for other methods and highlights the advantage of CESME in these settings.

\section{Conclusions and Discussions}

In this paper, we introduce a novel semiparametric mixture model that forms the basis of CESME, a new high-dimensional clustering method, which enjoys both strong empirical performance, outperforming widely used pipelines for scRNA-seq analysis, and rigorous theoretical guarantees. Statistically, effective clustering naturally aligns with a mixture model, making model-based clustering a principled approach to the problem. The success of CESME reinforces this perspective.  While CESME currently employs an EM-type alternating maximization algorithm, exploring more efficient optimization strategies remains an important direction for future work. The computational issue becomes even more critical when we extend the method to high-dimensional tensor data clustering. Recently, \cite{mai2021doubly} developed DEEM based on a mixture of tensor normals for high-dimensional clustering. It is natural to consider the tensor generalization of CESME, which will be reported in a separate paper.

The online supplementary file provides the following materials: (A) proofs of Theorem \ref{cor:upper_bound} with necessary technical lemmas, (B) discussions of identifiability of the model, (C) proofs of technical lemmas to establish Theorem \ref{cor:upper_bound}, (D) more technical lemmas and discussions on the initializations, (E) additional numerical results for the simulation study and real data analysis, (F) simulations of one-step EM variant of CESME and performance of CESME under mis-specified models.

\section*{Data Availability}
The datasets analyzed in this study are publicly available from previously published studies and are described in Section~\ref{sec:real}.
{\tt R} code for reproducing the numerical studies and data examples can be found at \url{https://github.com/lyuouzhang/CESME}.

\bibliographystyle{apalike}

\end{document}